\documentclass[aps,pra,twocolumn,showpacs,superscriptaddress, nofootinbib]{revtex4-2}

\usepackage[english]{babel}
\usepackage[utf8]{inputenc}

\usepackage[pdftex, pdftitle={Article}, pdfauthor={Author}]{hyperref} 

\usepackage{natbib}
\usepackage{hyperref}
\usepackage{verbatim}
\usepackage{amsmath}
\usepackage{amsfonts}
\usepackage{amssymb}
\usepackage{amsthm}
\usepackage{mathtools}
\usepackage{bm}
\usepackage{xparse}
\usepackage{graphicx}
\usepackage{xcolor}
\usepackage{textcase}
\usepackage{url}
\usepackage{lipsum}
\usepackage{appendix}
\usepackage{epstopdf}

\def\Tr{\mathrm{Tr}}

\newcommand{\diagdots}[3][-25]{%
  \rotatebox{#1}{\makebox[0pt]{\makebox[#2]{\xleaders\hbox{$\cdot$\hskip#3}\hfill\kern0pt}}}%
}

\DeclareDocumentCommand{\Tr}{m m O{\big}}{{\rm Tr}_{\:\!{#1}}#3({#2}#3)}

\begin{document}
\title{On playing gods: The fallacy of the many-worlds interpretation}

\author{Luis C.\ Barbado}
\affiliation{Institute for Quantum Optics and Quantum Information (IQOQI),
Austrian Academy of Sciences, Boltzmanngasse 3, A-1090 Vienna, Austria.}
\affiliation{Faculty of Physics, University of Vienna, Boltzmanngasse 5, A-1090 Vienna, Austria}

\author{Flavio Del Santo}
\affiliation{Group of Applied Physics, University of Geneva, 1211 Geneva 4, Switzerland; and  Constructor University, Geneva, Switzerland }

\date{\today}

\begin{abstract}
\noindent We present a methodological argument to refute the so-called \textit{many-worlds interpretation} (MWI) of quantum theory. Several known criticisms in the literature have already pointed out problematic aspects of this interpretation, such as the lack of a satisfactory account of probabilities, or the  huge ontological cost of MWI. Our criticism, however, does not go into the technical details of any version of MWI, but is at the same time more general and more radical. We show, in fact, that a whole class of theories--of which MWI is a prime example--fails to satisfy some basic tenets of science which we call \textit{facts about natural science.} The problem of approaches the likes of MWI is that, in order to reproduce the observed empirical evidence about any concrete quantum measurement outcome, they require as a tacit assumption that the theory \emph{does in fact} apply to an arbitrarily large range of phenomena, and ultimately to \textit{all} phenomena. We call this fallacy the \textit{holistic inference loop}, and we show that this is incompatible with the facts about natural science, rendering MWI untenable and dooming it to be refuted.

\end{abstract}
\maketitle

\section{Introduction}

Quantum mechanics is about to celebrate its first centenary and yet, as D.~Mermin pointed out, the “disagreement about the meaning of the theory is stronger than ever. New interpretations appear every day. None ever disappear” \cite{mermin2012commentary}. 

One of the most popular interpretations of quantum theory seems today to be the so-called \textit{many-worlds interpretation} (MWI) \cite{iii1973theory, everett1957relative,deutsch1985quantum, dewitt2015many, sep-qm-manyworlds, osnaghi2009origin, 10.1093/acprof:oso/9780199560561.001.0001, wallace2012emergent}. This interpretation supposedly solves the measurement problem by refuting the fact that quantum experiments yield single outcomes (among mutually exclusive possibilities), but rather all results are obtained in different worlds. John Bell pointed out some sociological origin in the rising consensus of this exotic interpretation: ``The MWI [\ldots] seems to attract especially quantum cosmologists, who wish to consider the world as a whole, and as a single quantum system, and so are particularly embarrassed by the requirement, in the pragmatic approach, for a `classical' part outside the quantum system\ldots\ i.e., outside the world." (\cite{bell2004speakable}, p. 193.). 

The MWI has been already criticized on several different accounts \cite{kent2010one, 10.1093/acprof:oso/9780199560561.003.0005,  Gisin2013, dawid2015many, BARRETT201731, 10.1093/acprof:oso/9780199560561.003.0006, gisin2022multiverse, dawid2022epistemic, 10.1093/acprof:oso/9780199560561.001.0001, kent2015does, pittphilsci21841}, especially for the ontological cost of its enormous inflation of worlds \cite{del2023potentiality}, the need of a preferred basis \cite{hemmo2022preferred, dawid2015many}, and its inability to account in a satisfactory way for the probabilistic nature of quantum predictions \cite{kent2010one, dawid2022epistemic, kent2015does} (see also Sect.~\ref{mwi}). In this paper we put forward a criticism that is at the same time more general (in fact, it applies to a whole class of theories of which the MWI is only an example) and more radical. Indeed, we will show that adopting an approach like the one propounded by the MWI leads to a fallacy that we call the \textit{holistic inference loop} (HIL). We will show that, in turn, HIL is incompatible with some basic \textit{facts about how natural science,} thereby rendering theories which enter the HIL untenable.


In a nutshell, we show that, to connect the MWI ontology with the observed measurement outcomes, it is necessary to assume that the theory is indeed strictly valid for an undefined, arbitrary set of phenomena, and ultimately for all phenomena in nature, a statement which can never be empirically proven. We notice critically how such statement on the universal validity of the theory is not taken just as a ``not-yet-falsified hypothesis'', for which certain limitations may be empirically found in the future. Rather, the statement that the theory \emph{does hold universally} constitutes a necessary assumption in the explanation of each an every single empirical verification of the theory. This sort of reasoning is what we identify with the aforementioned HIL.

Our arguments lead us to label MWI as an untenable physical theory. Such a sharp conclusion is a natural consequence of our criticism going to the roots of the methodological justification of the theory, unlike other criticisms that focus on specific elements within MWI. Therefore, if our arguments are accepted there is no possible redemption for MWI with future amendments: MWI should be rejected once and for all.


\section{Natural science as a human endeavour}
\label{human}

\subsection{Facts about natural sciences}
\label{facts}

We begin by highlighting three quite evident, but important facts about science, which are a direct consequence of it being a purely human activity. Due to the hardly debatable nature of the following three statements about science, we will simply refer to them as \emph{facts about natural science:}

\begin{itemize}

    \item[(i)] \emph{Empirical data are limited by the range of observations.}\\
The empirical data at our disposal to test our theories are ultimately obtained from the observation of natural phenomena,\footnote{We are not here taking any radical antirealistic stance by claiming that \emph{only} sensorial experiences are meaningful.
We merely claim that, as a matter of fact, all the available data in science are acquired through observation, i.e., through human senses, often aided by instruments purposely built, in turn, by humans.} and hence they are obviously limited in several different ways. Among the many limitations, we focus on the necessarily limited range of natural phenomena covered by human observations. No matter how much we strive for enlarging such range, at any point in time there will always be phenomena that lie beyond current experimental reach. 

    \item[(ii)] \emph{Scientific theories are human creations.}\\
The scientific theories that we use to organize our data and conceptually explain (our sensorial experience of) the natural phenomena are purely a human construction. It does not matter how astonishingly well a theory seems to fit with a huge range of empirical evidence, or even with all evidence at disposal at a certain stage of the development of science, there is always a fundamental distinction between the concepts of the theory and our available empirical data. The former may \emph{match} the latter, but they  \emph{are not} the latter.\footnote{A similar position to the content of (i) and (ii) was upheld by D.~Bohm, perhaps even more radically because he gave to the limitations a more ontological character: ``Any given set of qualities and properties of matter and categories of laws that are expressed in terms of these qualities and properties is in general applicable only within limited contexts, over limited ranges of conditions and to limited degrees of approximation, these limits being subject to better and better determination with the aid of further scientific research. Indeed, both the very character of the empirical data and the results of a more detailed logical analysis show that beyond the above limitations on the validity of any given theory, the possibility is always open that there may exist an unlimited variety of additional properties, qualities, entities, systems, levels, etc., to which apply correspondingly new kinds of laws of nature.'' \cite{bohm1957causality}, p. 133.}

    \item[(iii)]  \emph{Valid scientific theories are self-consistent and empirically adequate.}\\
Necessary conditions for the validity of a theory in natural science are its internal consistency \emph{and} its empirical adequacy--namely, the degree to which it fits and predicts empirical data--\emph{within} its purposed range of applicability.\footnote{Clearly empirical adequacy is endorsed by both realists (who take this as a sign that a theory is approaching truth) and empiricists (for whom this is the main, and perhaps the only necessary, standard of science); see, e.g., \cite{bhakthavatsalam2017s}.} 
Within natural science, a theory should be devoted, besides yielding predictions, to provide satisfactory explanations for the empirical observations. This means to be able to formulate consistent stories that account for the observed phenomena. Note that this may involve introducing into a theory metaphysical and mathematical elements (such as particles, forces, fields, quantum states, etc.), as well as correspondence rules between the observed data and these terms (see, e.g., \cite{jammer1974philosophy}).\footnote{\label{note}A disclaimer is here in order. In quantum foundation it is customary to distinguish between theories and interpretations.  What we mean here by theory encompasses not only the empirical content and the mathematical structure, but also the whole collection of postulated metaphysical entities. Therefore, in this paper we will use theory and interpretation interchangeably unless differently specified.}

\end{itemize}

Without commitments regarding metaphysical positions, we claim that the previous facts about natural science are compatible with a wide variety of views about science. In fact, independently of whether one has a strong realistic or a more empiricist view (or even a solipsistic view, for that matter), who is in the position to refute that we only have  limited empirical data and that we will never have a “complete” collection thereof? Does anyone have a collection of empirical data that covers all natural phenomena we can ever experience? Does anyone visualize a concrete plan that will provide us at some point with such a collection? The same applies for the human nature of scientific theories and concepts: Who can refute that any scientific theory we know is a creation of humankind? As a matter of fact, it was not flies, dogs, or monkeys that created scientific theories, but humans did, in the same way that they dominated fire, developed natural languages, invented religions, or built the Pyramids. 
Scientific theories do not get revealed to us in any Tablets of Stone: All we have done and can do, \emph{it is a fact,} is to create our own theories and test them against our \emph{limited} empirical evidence.
Finally, the fact that the validity of a theory is given by its capacity to reproduce the empirical evidence is simply a desideratum of natural science by definition, to distinguish it from different kinds of knowledge.


\subsection{Tenability of scientific theories}\label{consequences}

Having in mind the facts about natural science that we have stated, let us now go into the philosophical discussion on what implications these facts have when critically analyzing any scientific theory. In this sense, we propose the following elementary statement:

\vspace{0.5cm}

\noindent \emph{\textbf{Necessary condition of tenability (NCT) for a scientific theory:} Any theory that enters in contradiction with the facts about natural science, should be deemed untenable.}\footnote{For our argument, we do not need to engage into the notorious debate on the problem of induction and demarcation (see~\cite{popper2005logic}), either based on verifiability or falsifiability. We merely take the compliance of a theory with the facts about natural science as a minimal necessary condition of tenability.}

\vspace{0.5cm}
As radical as it may sound, we believe that it is a completely fair requisite for any scientific theory. Any theory whose validity relies on assumptions that contradict the evidence about how science is indeed constructed is untenable. This reasoning is based on a descriptive argument: It would be a theory that is \emph{pretending to do something that it is not doing.} 
 
A theory may enter in conflict with the facts about natural science in many ways. We wish to bring here one specific way, which we shall call the \emph{holistic inference loop,} on which we will base our criticism of  the many-worlds interpretation (Sect.~\ref{mwi}).
Let us call some scientific theory~$T$, some specific range of known phenomena~$P$ that we wish to scientifically explain with said theory,\footnote{$P$ needs not to be the set of \textit{all} phenomena that~$T$ aims to account for. It could be also a subset thereof, but in any case it must be of course limited, in the sense discussed in fact (i). Even in the case that~$T$ aims to be an ``universal theory'', in whatever sense, it is always only tested against the set of all experienced phenomena, which is necessarily limited.} and the empirical data we have collected about such phenomena~$E_P$. Then one can formulate the following:

\vspace{0.5cm}

    \noindent \emph{\textbf{Holistic inference loop (HIL):} A theory~$T$ enters the HIL if the inferential connection between~$E_P$ and~$T$, which allegedly justifies the validity of~$T$ for~$P$, necessarily relies on the assumption that~$T$ applies to some other arbitrarily large range of phenomena~$P’$.}

\vspace{0.5cm}

We contend that 

\vspace{0.5cm}

\noindent \emph{\textbf{HIL implies untenability:} A theory~$T$ that enters the HIL cannot meet the~NCT, and should therefore be rejected.}

\vspace{0.5cm}

 Indeed, meeting the~NCT means that~$T$ is compatible with the simultaneous fulfilment of the three facts about natural sciences~(i), (ii) and~(iii) (Subsect.~\ref{facts}). According to~(iii), for a theory $T$  to explain the set of phenomena~$P$ means to be able to fit and predict the available observed data~$E_P$. However, if a theory enters the~HIL, such capacity is conditioned on the fulfilment of a tacit hypothesis, which we shall call the \emph{holistic hypothesis,} namely that~$T$ also applies to other \textit{arbitrarily large} set of phenomena~$P'$. In other words, $T$ explains $P$ \textit{only if} it applies to an arbitrarily set of phenomena~$P'$. However, in agreement with~(ii), we understand the contingent nature of the theory~$T$ as a human creation, and this implies that we cannot assume its applicability to some arbitrary phenomena~$P'$ as granted. One can only assume this as a provisional hypothesis, but then the only possibility to check the hypothesis would be, again according to~(iii), to empirically show that~$T$ does indeed apply to~$P'$. According to~(i), however, this is not possible due to the arbitrary nature of~$P'$: We can never empirically exhaust an arbitrarily large set of phenomena. Therefore, the connection between our theory~$T$ and the empirical data~$E_P$ not only remains dependent on a hypothesis, but rather on a hypothesis that is impossible to corroborate while fulfilling the~NCT! It is therefore also impossible to justify the theory~$T$ from~$E_P$, and thus according to~(iii) the theory~$T$ has not been validated. In conclusion, if~$T$ enters the~HIL we cannot justify its validity for~$P$ while fulfilling the~NCT, and~$T$ is therefore an untenable theory.

\begin{figure}
    \centering
    \includegraphics[width=1\linewidth]{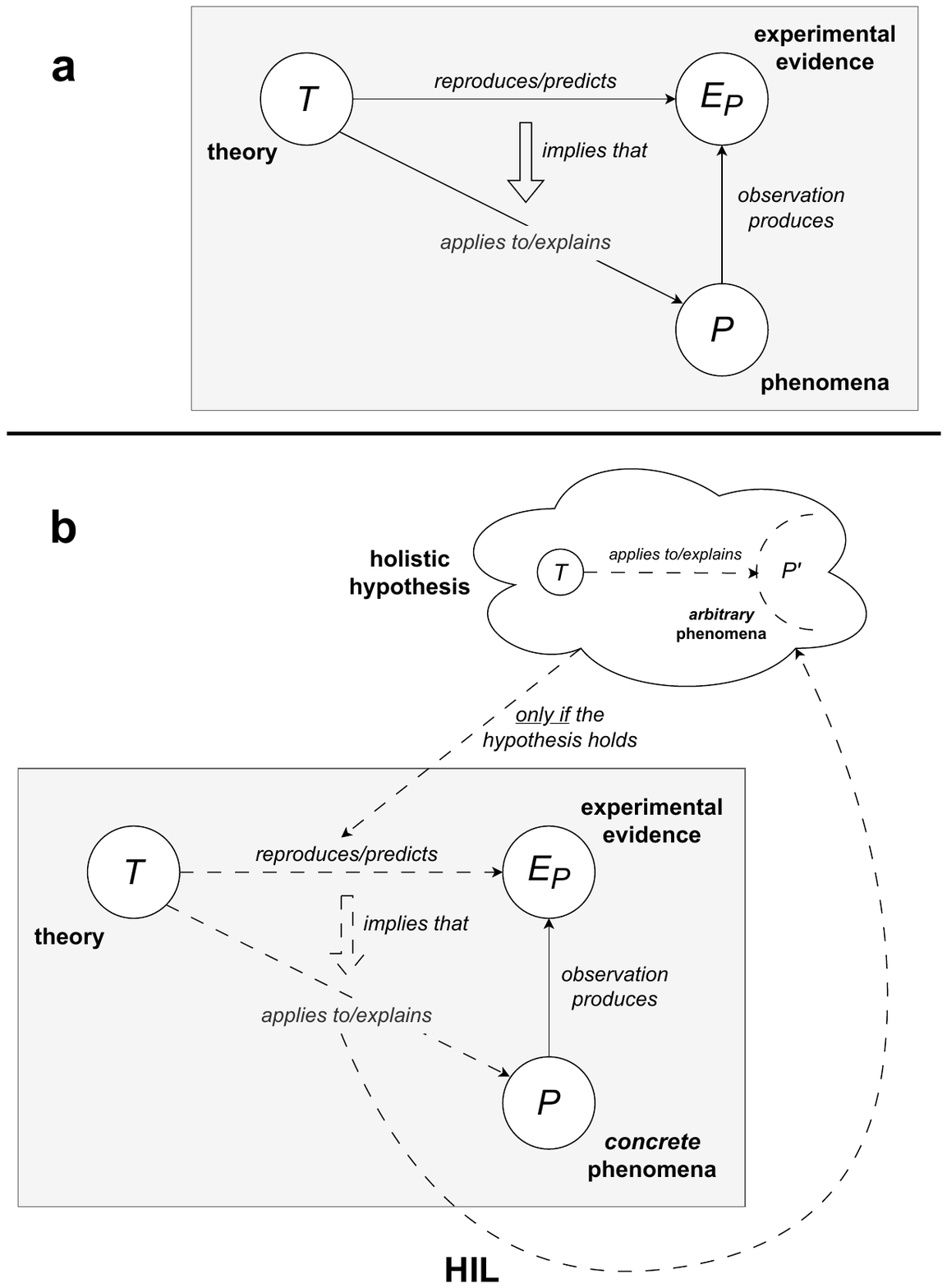}
    \caption{Schematic depiction of a) a standard validation procedure of the theory~$T$, and b) an untenable validation of the theory~$T$ that enters the holistic inference loop.}
    \label{scheme}
\end{figure}

In Fig.~\ref{scheme} we show in a) a correct validation of the theory~$T$ as applicable to the phenomena~$P$, due to its capacity to reproduce and predict the empirical evidence~$E_P$ of such phenomena; and in b) the scheme for a theory~$T$ entering the HIL. Let us pay attention to the two crucial aspects of such scheme:

\begin{enumerate}
    \item[I.] The range of phenomena~$P'$ within the holistic hypothesis must be an \emph{arbitrary} range. If~$P'$ is a well-defined finite range of phenomena, the theory is not entering the HIL. The reason is that, in such case, the validity of~$T$ for~$P'$, and with it the validity of~$T$ for~$P$, could be proven while fulfilling the~NCT. Usually, one would simply be in a \emph{legitimate} holistic scenario, i.e.,\ a scenario where the applicability of~$T$ to several phenomena~$P,~P',\ldots$ (provided that the total range remains well-defined and finite) cannot be understood separately, but rather \emph{altogether.}\footnote{Paradigmatic examples of holistic scenarios appear for example in ecology. Consider that we are studying whether a theory~$T$ applies to the feeding behaviour~$P$ of a certain species of predator in a given ecosystem. In order to do so, we make use of some collected empirical data~$E_P$ about the ecosystem. However, since all the different elements in an ecosystem are interconnected, $E_P$ is always highly dependent on many other phenomena in the ecosystem. So we may also decide to assume certain hypothesis about other phenomena~$P'$ in it; for example, we may assume that the theory~$T$ also applies to the feeding behaviour of other competing predators, \emph{even when we do not have full evidence for it.} If we prove this way that~$T$ applies to~$P$, the validity of such conclusion will be conditioned to out hypothesis that~$T$ applies to~$P'$. However, this is a well-defined hypothesis that could be proven right or wrong with further empirical research.}
    
    \item[II.] The holistic hypothesis must systematically appear as a necessary condition to invoke when justifying the theory~$T$ from the empirical data~$E_P$. This is perhaps the most distinctive aspect of the theories entering the~HIL, so let us stress the point: \emph{The formulation of a holistic hypothesis alone is not sufficient for entering the HIL.} The hypothesis must also constitute an omnipresent necessary argument any time one wishes to experimentally justify the theory for some concrete phenomena.\footnote{Classical mechanics is a clear example of a theory that presumes an arbitrary applicability and yet does not enter the HIL. We deepen on this idea in Subsect.~\ref{tales}.}

    
\end{enumerate}

Finally, let us clarify the meaning of the curve that closes the loop in~Fig.~\ref{scheme}, which we actually labelled `HIL' in the diagram. 
This represents the ``psychological'' fallacy that leads to propose the holistic hypothesis in the first place:
Namely, the belief that there is already strong enough empirical evidence supporting~$T$ (as applicable to certain phenomena~$P$) as to suggest the extrapolation of its validity to an arbitrary set of phenomena~$P'$; that is, as to formulate the holistic hypothesis. However, as we already argued, there is no empirical evidence that can ever fully validate the holistic hypothesis.
Such hypothesis will either be falsified or remain  empirically unsupported.
But then, we stress again: This is \emph{not} the full problem of the HIL. The \emph{full} problem is that the whole theory~$T$, including any concrete explanations of experienced phenomena that ``inspired'' the holistic hypothesis, relies on such hypothesis being strictly true. The circular argument is manifest! Hence, the whole theory~$T$ remains as devoid of a validation as the holistic hypothesis itself. 

The construction of theories that do not fulfil the~NCT leads, as an inevitable consequence, to fatal consistency problems in such theories and in their relation with the empirical evidence. In particular, for the theories entering the~HIL we identify major problems both for the robustness of their statements and, at the same time, for the prospects to overcome these theories with better theories in the future. We will further discuss these problems in details through the concrete case of the MWI, which we address in the next section as a paradigmatic example of a theory that enters the HIL.

\section{The Many-worlds Interpretation}
\label{mwi}

\subsection{Interlude: The empirical evidence for quantum mechanics}

At the level of empirical evidence, what we find for whatever range of quantum phenomena~$P$ is: 1) That each time a quantum measurement takes place a single outcome is found, 2) that for a measurement repeated under empirically identical conditions, the outcomes of each repetition are in general different, and their long run statistics match the probabilities predicted by the Born rule for the given conditions, and 3) that the relation between the statistics of successive measurements for isolated physical systems can always be reproduced by the unitary evolution given by the Schr\"odinger equation. These are the three (interrelated) parts in which we can organize the set of available empirical evidence~$E_P$ for quantum mechanics. Any theory or interpretation~$T$ of quantum mechanics must, first and foremost, reproduce and provide an explanation for \emph{all three} parts, at least to the accuracy to which they have been tested.

\subsection{How MWI falls into the HIL} \label{MWI_unt}

The Many-worlds interpretation (MWI) was introduced by H.~Everett in 1957 (initially called relative-state interpretation) \cite{everett1957relative, osnaghi2009origin}. It maintains  that ``every time a quantum experiment with different possible outcomes is performed, all outcomes are obtained, each in a different newly created world'' \cite{sep-qm-manyworlds}. As we mentioned, the MWI has received a profuse collection of criticisms \cite{kent2010one, 10.1093/acprof:oso/9780199560561.003.0005,  Gisin2013, dawid2015many, BARRETT201731, 10.1093/acprof:oso/9780199560561.003.0006, gisin2022multiverse}. Most of these can be broadly simplified into questions, such as: Where are all those worlds supposed to live? When, where and under what circumstances the branching happens? Which is the preferred basis in which such branching occurs? \cite{dawid2015many}. And, most of all, how do the (observed) probabilities arise from the unitary (deterministic) dynamics? \cite{dawid2022epistemic, wallace2012emergent, 10.1093/acprof:oso/9780199560561.001.0001, kent2015does, wilson2013objective}. Although we share these critical views, the argument against MWI we propose here, as we already advanced, does not address any of these 
problems within the theory. So, in what follows, we can as well assume that these known criticisms can be somehow overcome.

Let us show how the MWI enters the HIL. This theory, in fact, elects the unitary evolution as \emph{the} fundamental law of nature, which applies to literally everything, i.e., to the whole state of the Universe. So it automatically reproduces, just by definition, the part of the empirical evidence that corresponds to the relations between statistic outcomes of different measurements, given by the proposed unitary evolution. But what about the fact that only one outcome is empirically found each time a measurement takes place, and about the statistics of those outcomes themselves? In this case, the reasoning that connects these empirical evidences~$E_P$, for whatever quantum measurement~$P$ we wished to describe, with 
the theory~$T$, \textit{necessarily assumes} that all such quantum mechanical concepts not only satisfactorily describe~$P$, but also other phenomena~$P'$. The latter include the evolution of the measurement apparatuses, our brains, consciousness, and ultimately the entire Universe. This can be explicitly found in the words of Everett himself~\cite{iii1973theory}: ``[A]ssume the universal validity of the quantum description, by the complete abandonment of [the Born rule plus the state-update rule]. The general validity of pure wave mechanics, \emph{without any statistical assertions}, is assumed for all physical systems, including observers and measuring apparata. [\ldots] [This alternative interpretation] has the virtue of logical simplicity and it is complete in the sense that it is applicable to the entire universe.'' The HIL is thus here striking because, in order for the theory~$T$ to account for any specific~$E_P$ in a consistent way, one needs to assume the universal applicability of the theory, i.e., $P'$ must be the whole Universe.

We emphasize the critical point: The problem is not only the assumption that MWI applies universally, the problem is the use, or rather abuse, made of that statement. Other well-established physical theories also make formal claims on their universal applicability. However, when contrasting those theories with the experimental evidence, one does not need to invoke such universality as something strictly true. Rather, one only requires that the theory applies to the finite range of phenomena which is significantly involved in the specific experiment under consideration. On the contrary, for MWI strict universality is a tacit necessary condition appearing in each and every single account of whatever quantum experiment.

For the self-consistency of MWI, universality must strictly mean that the theory does apply to \emph{any natural phenomena that we may ever access empirically,} which is definitely an arbitrary range of phenomena~$P'$. One could at first glance hesitate whether this is really a strict necessity of the theory to account for each concrete measurement outcome we perceive. After all, consider that we rephrase our empirical evidence on ``the outcomes of measurements'' as, say, ``the perception by our consciousness of the outcomes of measurements''. Would it then be enough if we could just prove that ``our consciousness'' (if we ever manage to define it accurately enough) evolves unitarily according to quantum mechanics? We argue that this is not the case, by simply noticing that any phenomena~$P'$ we may ever access in nature must possibly become affected by results of quantum measurement outcomes just as our consciousness is. Therefore, any depart from quantum mechanics by~$P'$ would equally compromise the MWI account, not only of~$P'$, but of the result of the quantum measurement. In other words, any possible phenomena we may ever access empirically can play the role of the observer in a quantum measurement, in the sense of becoming affected by its result. Therefore, if the requisite that \emph{any} observer is governed by quantum mechanics is a necessary ingredient in the understanding of the measurement process, and of the consistency of the empirical evidences on the individual outcomes of it, then the validity of such understanding becomes conditioned to quantum mechanics governing \textit{all} phenomena.

For example, in a more recent work \cite{quantum5020031} on a variant of the MWI, the following is stated on the perception of the trajectory of an alpha-particle emitted in a cloud chamber: ``The reason for the fact that we only see one trajectory at a time is that even though everything is a q[uantum]-wave within which things exist at the same time, when we interact with this q-wave we can only reveal some of its aspects, one at a time (this is where Heisenberg’s Uncertainty comes from). We ourselves are also a collection of q-waves and it is when our q-waves correlate with the q-waves of the alpha-particle that c[lassical]-numbers emerge. These correlations between q-waves are called quantum entanglement and so the classical world owes its own existence to quantum entanglement.'' Here we find explicitly how the experimental evidence on the alpha-particle trajectory can only be accounted for due to the quantum nature of the observer. But then, either strictly \emph{everything} in nature is truly a quantum wave, or making something which is not to play the role of the observer would ruin the given MWI explanation, also for the alpha-particle and for our perception of it. We will explore this idea in further detail, and its fatal consequences for the MWI, in the next subsection.

The analysis of MWI as a theory entering the HIL reveals as glaring the circular, self-justified nature of its supposed validity. Indeed, without the \emph{a priori} assumption that any phenomena can be accounted for with an unitary evolution, the supporters of many-worlds have absolutely no way to justify the empirical adequacy of quantum mechanics, because they cannot reproduce the statistic outcomes of measurements. Following the psychological fallacy of the HIL that we mentioned in the previous section, in order to elevate quantum mechanics to its universal status, MWI supporters resort to the fact that there is an outstanding empirical evidence for quantum mechanics, indeed the best ever achieved for any physical theory. But such empirical adequacy of quantum theory, i.e.\ the success of its empirical predictions, was allowed by testing it against theories that account for probabilistic outcomes at the fundamental level, predominantly the Copenhagen Interpretation. These are all theories that consider a transition between the quantum and the classical domain, i.e., that display a Heisenberg cut.\footnote{These theories can be divided into two main families: those that take the cut to be epistemic and arbitrarily movable (such as more or less subjective variants of the Copenhagen interpretation \cite{laudisa2002relational, fuchs2010qbism, auffeves2016contexts, brukner2017quantum, drossel2018contextual}), or the objective collapse models (\cite{ghirardi1986unified, gisin1989stochastic, ghirardi1990markov}).} These theories manage to separately account for the empirical evidence of each limited quantum phenomena one may wish to explore, without needing to tacitly involve the whole Universe in each an every experiment. The accumulation of an overwhelming amount of successes of those individual experiments, the self-consistency among them, and the lack of any failure to date, may suggest distinguishing the theory with some honour of ``universality'' (as of today). But such honour is only a late award for a flawless career of well-established and coherent achievements. In contrast, for MWI the medal of being a \emph{strictly universal} theory is the initial (unaffordable) deposit that the theory demands in advance, before gathering its very first successful empirical prediction.

We propose the MWI supporters the following simple thought experiment: Convince a classical physicist, who knows nothing about quantum mechanics, of the validity of MWI as accounting for the empirical evidence of some very simple quantum experiment. Notice that we say the validity \emph{of MWI} specifically. That is, the whole explanation should be done \emph{strictly without ever mentioning any notions of collapse, measurement outcome, probabilities, Heisenberg cut, etc.,} at least at a fundamental level. It is not difficult to realize how puzzled the classical physicist will be. But the astonishment will not come from a fascination about the existence of those unfathomable parallel universes. It will rather be much more worldly: She will need to accept that this completely new theory, of which she has not been given any single empirical evidence yet, does indeed apply, not only to the simple quantum system that she is about to test in the laboratory, but also to other arbitrary phenomena, including her own consciousness, \emph{as a pre-condition} to possibly make any connection between the theory and the empirical evidence that she will obtain from the experiment. That is, she needs \emph{first} to accept, as an \emph{act of faith,} such arbitrary range of applicability \emph{before} she gets her very first empirical verification of MWI in, say, a double-slit experiment. 
It is again not difficult to realize, following our previous discussion, that within the MWI there will be no empirical verification she will ever obtain that is not conditioned to further acts of faith.

We highlight what the previous discussion ultimately evinces: MWI can be considered strictly as an \emph{interpretation} of quantum mechanics, with identical empirical predictions to other interpretations, \emph{only if} one illegitimately assumes that universality does hold, whereas other interpretations do not require such assumption. We have thus revealed how the MWI rhetoric has hidden its own empirical void behind the enormous empirical success of \emph{the other} interpretations. The seemingly religious need of unification and simplification of ``many-worlders'' leads them to take the success of these theories, and add to them the unwarranted claim on the universal domain of validity of unitary evolution, to finally try to relate MWI with the empirical evidence about even the simplest quantum phenomenon: They need to play gods to justify their theory.

\subsection{Lack of robustness: The many-worlds interpretation is doomed to be refuted}\label{robustness}

As we already discussed, entering the~HIL leads to the untenability of a theory. Here we address a trivial consequence of it, namely the lack of robustness of the description of reality made by MWI.

Many-worlders claim that their theory is empirically testable because quantum mechanics is empirically testable. Let us assume that. The point is that, if the explanation of what happens with some \emph{concrete} phenomena~$P$ relies on the universality of the theory, then if one day we find \emph{whatever} phenomenon that does not fit into quantum mechanics, it means that we don't understand~$P$ with MWI any more. Actually, we suddenly don't understand anything at all. Our comprehension of the basic quantum phenomena such as the Stern-Gerlach experiment, the nucleus decay, the structure of the atom, etc.\ just falls apart!

Let us give some concrete example on how this would actually happen. Consider a Stern-Gerlach experiment in which, depending on the trajectory in which the spin particle is deflected by the magnetic field, a macroscopic device raises either a yellow or a green flag. The many-worlds description of the experiment is that the spin orientation, the particle trajectory and the colour of the raised flag become all entangled during the process. The fact that we perceive sometimes a yellow flag and sometimes a green flag is only a consequence of our own (quantum) consciousness further entangling with all those elements. But, in the many-worlds ontology, the reality of the flag is neither yellow nor green. The reality of the flag, as that of the rest of the elements in the experiment, is given by a wave function that describes a collection of worlds, in each of which the flag and the other elements have different properties.
New physical theories to come may conceptualize the reality of this experiment very differently, and eventually more accurately. But at least for experiments describable by quantum mechanics, such as this one, they \emph{must reproduce the existence of these many-worlds} as a very accurate and consistent description. This would be the desired robustness of the theory that we will prove to be nonexistent.

Now, a simple question can be made: Any time after the experiment took place, can we use the colour of the raised flag that we perceive as an empirical evidence for other experiments? Say, for example, that we wished to study the colour perception of certain families of insects, or we wished to test several brands of detergent on different colouring of clothes. Could we rely on the colour of the flag that we perceive and use it, for such purposes, as an \emph{empirical fact?} The answer within many-worlds, we believe, would be affirmative. But notice that this is done even when our perception of the flag and the reality of the flag are completely different things! The reason why this is possible is that our own perception is telling us that we are in the branch where the flag is, say, yellow; and \emph{noticing that all the further elements involved (the insects, the detergents\ldots) are also quantum mechanical,} a simple computation shows that the interference with the branch in which the flag is green is completely negligible. So, for practical purposes, in order to keep track of our own perception of the further experiments we want to carry, we can take the colour of the flag as an empirical fact, and everything remains consistent with the simultaneous existence of the rest of the wave function.

But then, what about testing a potentially post-quantum phenomena? Can we use the colour of the flag? Consider that we have found some new natural phenomenon, which we wish to study experimentally in order to check whether we can describe it with quantum mechanics or not.
Notice that, on the one hand, we already have a very accurate description of the flag by MWI, so we can quite safely state that the flag is ultimately neither yellow nor green. But, on the other hand, the phenomenon that will become affected by the reality of the flag may not be describable with MWI, so we cannot rely anymore in the fact that the overlapping between the realities with different colours will stay negligible. Hence, we shall not trust anymore our own perception of the colour and use it directly as an empirical evidence, since it is strictly a subjective perception which consistency is only warranted for other quantum systems. Maybe we need to be more careful and consider the full reality of the flag that MWI provides us, namely the wave function of it.

But then, if we cannot use the colour of the flag as an empirical evidence, a huge problem arises: Can we use \emph{any} outcome of a quantum measurement whatsoever as empirical evidence in order to test some potentially post-quantum phenomenon? We notice how the consequences of a negative answer to this question are completely unacceptable: All our sensorial experience, even the seemingly most stable one, can be described, to a great extent, as a consequence of quantum phenomena; and thus its reality is, at least to such extent, describable by MWI. Therefore, we can safely state that all reality around us, meaning the wave function of it, and our sensorial perception of it, are extremely different things. We cannot then rely on our sensorial experience at all in order to do experiments beyond the quantum regime, and it is thus completely impossible to empirically test any post-quantum phenomenon.\footnote{Any attempt to compute the ``really existing wave-function'' as the correct evidence to use is also hopeless: Any computation of any wave function must itself rely on the use of some other sensorial experiences that cannot be trusted themselves either, for the same reasons.}
Such conclusion is a natural consequence of MWI assuming the universality of quantum mechanics in order to make sense of our sensorial experience. But this is of course untenable, because it would mean that quantum mechanics cannot be refuted experimentally.

If we want to avoid such blind alley, we have to accept that at least some sensorial experiences can be considered as empirical evidences on their own right, that we could involve into experiments testing potentially post-quantum phenomena. For the sake of continuity, let us symbolically pick again the colour of our flag as one of those evidences. We then make some parameters of the potentially post-quantum experiments we are carrying to be dependent on such colour. Consider now that we find some post-quantum phenomenon, something which description does not fit into the quantum mechanics formalism. Of course, the results of those post-quantum experiments will be consistent with the colour of the flag. And notice that, by stating this, we are assuming absolutely nothing about such post-quantum phenomenon. Nothing except that we can use the colour of the flag as an empirical evidence.

But then, how is it that we find consistencies, even for the post-quantum phenomenon, with the flag being yellow when we see it yellow and with it being green when we see it green? Notice that the usual many-worlds explanation is now not warranted. We cannot say that our post-quantum system got entangled with the flag and that we got two copies of our system, one acting as if the flag was yellow and the other as if the flag was green. That would truly mean we are imposing properties to our post-quantum system that we have no right to impose at all. Ultimately, in order to reconcile the situation with the many-worlds reality of the flag that we were so sure about, we would need to impose that our post-quantum phenomenon is actually a quantum one!

Again, we encounter ourselves with the need to enforce universality of quantum mechanics so that the connection between the proposed realities \emph{of the quantum flag, particle trajectory and spin,} and the empirical evidence we have about them can survive at all.
In conclusion, as we already advanced, finding just a single post-quantum phenomenon, we would find out that MWI describes absolutely no phenomenon of the reality we live in, in no specific limit and to no reliable degree of accuracy.\footnote{One may dismiss our argument as an illegitimate criticism to MWI, as soon as it involves (potentially) post-quantum phenomena to which MWI does not pretend to apply. However, it is not us who make the nefarious decision of involving those post-quantum phenomena. It is MWI itself that does so, as soon as its explanation of the outcomes of quantum measurements makes statements about ``the observer'' of such outcomes. Given such explanation, we just need to further notice that post-quantum phenomena may exist and may play the role of that observer, and tragedy unfolds on its own. As importantly, we remark that our argument is \emph{not} requesting MWI to provide an \emph{a priori} recipe on how it should be understood in coexistence with the next theory to come. We are just pointing how it fails to account for the empirical evidence of a quantum measurement outcome without assuming that whatever else that coexists with such outcome must also be quantum.}

\begin{figure}
    \centering
    \includegraphics[width=1\linewidth]{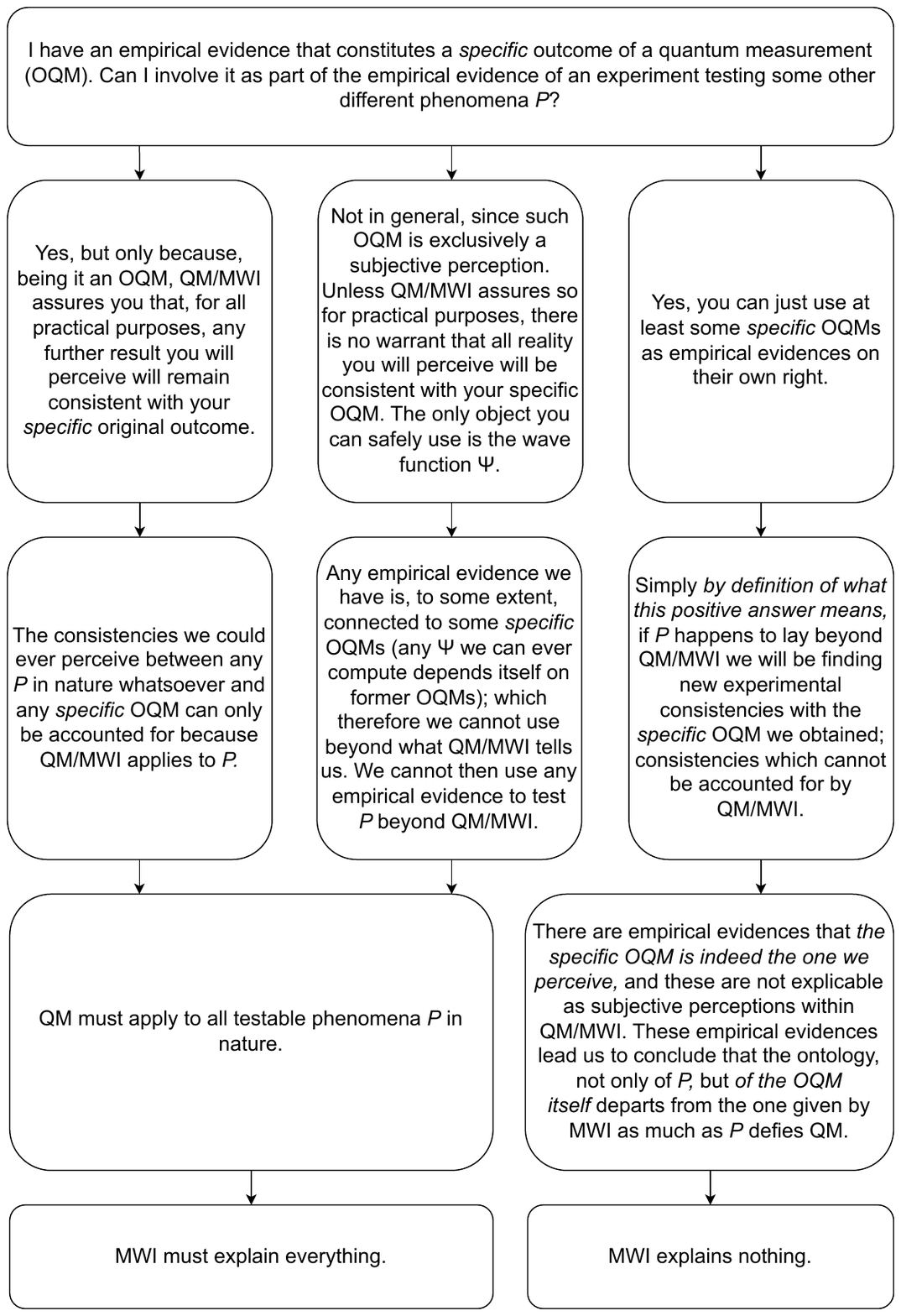}
    \caption{Scheme with possible answers on whether to consider the \emph{specific} outcome of a quantum measurement as an empirical evidence for other phenomena. `QM' stands for \emph{quantum mechanics.}}
    \label{question}
\end{figure}

The given example is just a detailed staging of how entering the~HIL compromises the robustness of the statements about reality that any theory can make. In Fig.~\ref{question}, we provide the reasoning scheme of the example presented in a generic way. While following it, one easily realizes how the problem arising is always the same: The need to be consistent with an unjustified assumption of universality, so that the full house of cards does not crumble. Therefore, beyond the detailed follow-up of the reasoning, our core argument is precise and simple: Making a theory completely dependent on its universality to actually hold for anything concrete whatsoever is an \emph{all in} bet: Either one ``explains everything'', or nothing at all (see Fig.~\ref{question}). But such bet in natural sciences is doomed to lose, sooner or later, due to our always limited experimental evidence, and the always existing possibility that new phenomena defying the theory shows up. Therefore, MWI could at most survive as a mental scheme in a ``pure-quantum universe'' of fantasy, but not as a valid theory of any concrete phenomena in the natural world we happen to live in.\footnote{We bring at this point the following quote from the approach to MWI in~\cite{deutsch1985quantum}: ``Whether or not a theory is universal is a purely logical property of the theory. This must be distinguished from the empirical question whether quantum theory is universally true. The latter question is not addressed in this paper.'' It seems thus that some approaches to MWI are indeed explicitly dedicated to build up (on paper) a self-sufficient and self-consistent pure-quantum universe. Put like that, there is nothing really problematic in such formal endeavour. This article simply shows how, if the self-consistency is so fragile that it does not admit any interaction with a non-quantum ingredient without fully breaking (equivalently, it presumes universality of quantum mechanics as an omnipresent internal requisite), the price to pay for building such beautiful universe is to strictly detach it from any tenable description of any natural phenomena.}

\subsection{Interlude: On refutations and overcomings}\label{tales}

In this subsection we further stress that, once a single empirical evidence escaping a theory that enters the HIL shows up, we ought to \emph{complete refute} it. The theory is not simply overcome by a more general one that encompasses the previous theory as a limiting case.

As a paradigmatic example of a fully refuted theory, although for purely experimental reasons, let us consider phlogiston theory. The different formulations of this theory could apparently account for several aspects of the combustion and rusting processes. However, all these supposed explanations relied on an unproven hypothesis: The existence of a substance, namely \emph{phlogiston.} When further experiments and more rigorous analyses showed evidences that were incompatible with such substance, the theory was fully refuted: All the supposed explanations made by the theory were \emph{strictly wrong.} They became fairy tales.


This fate is in blatant contrast with, for example, that of classical mechanics and Newton's gravity. We still can understand and very accurately explain the Solar System with their old concepts, regardless of quantum mechanics, special and general relativity, or any broader theory that we will find tomorrow. Indeed, all these new theories, far from fully refuting the old ones, only reaffirm their validity within their range of applicability: Rather than strictly wrong, they are \emph{approximately correct} theories. Notice in particular, that it is not just specific results of the old theories that are recovered: It is \emph{the full theories} that one can find in the appropriate limits. Therefore, the old theories were \emph{overcome,} but \emph{not} fully refuted.


The reason that those theories survived as approximately correct is clear: Their account of the empirical evidence did not depend on an unwarranted hypothesis. In particular, and coming back to what matters for contrasting them with theories entering the HIL: In describing, say, the Solar System, the arguments connecting the theories with the empirical evidence do not make the claim that ``our consciousness'', or any \emph{arbitrary} phenomena that may be affected by the motion of the planets, need to behave according to classical laws or concepts.

We notice how some of the principles in classical mechanics may receive the epithet of ``universal'', famously Newton's law of gravitation. But this is because they are such \emph{within} the theory. However, there is no need to invoke that this universality is strictly true in nature every time one applies the principle to a concrete situation (indeed, we know today that Newton's specific formulation is not universal). Something similar can be said about the concepts of absolute time or Euclidean space, which are assumed to ``apply universaly''. If we are describing what is happening on a billiard table, it is enough for the theory if these concepts apply (very accurately) in that regime, something of which we have astounding empirical evidence. Given that \emph{proven} assumption, the theory already manages to reproduce the empirical evidence on the motion of the billiard balls, and its duties are accomplished. We refer to Subsect.~\ref{consequences} to easily realize why, in view of these facts, classical mechanics does \emph{not} enter the HIL.



An equally honorable future awaits the interpretations of quantum theory (and here we mean \emph{interpretations} in the strict sense, see footnote \ref{note}) that are provided with a Heisenberg cut. The latter limits itself to state that, in certain contexts, one evolution (the unitary one) is more suitable than the other (the ``collapse''), without explaining why this is the case. And such a statement, even if we may feel it incomplete or not fully satisfactory (for a variety of reasons), will nonetheless remain approximately true and useful, no matter what possible future theory or interpretation will revolutionize the worldview of quantum mechanics one day.

On the contrary, with the assumption of MWI as the correct interpretation of quantum mechanics, similarly to phlogiston theory, each and every supposed empirical verification is yet pending on an unproven hypothesis: That the theory is indeed strictly universal. And, worse off than with phlogiston theory, this hypothesis is not only unproven, but also impossible to prove! Therefore, the arguments exposed in this article show that the fate of MWI is inescapably that of phlogiston theory: To be fully refuted.\footnote{The rejection of MWI is frequently compared with former na\"{\i}ve scepticism about currently well-established theories (within their range of applicability), based on an ingenuous embracing of our most immediate perception of reality, such as scepticism about atomic theory~\cite{kaku} or even about Copernicus' heliocentric model~\cite{everett1957relative}. We hope that the arguments exposed here make it clear, without need of further discussion, how such comparisons are blatantly fallacious. We simply suggest, as an amusing counterfactual story to think about, what would have become today of Copernicus' model if he had grounded it on whatever deep statements about, say, the \emph{ultimate} functioning of our brain and the emergence of our consciousness that he could manage to propose in the XVI century.}


%

\subsection{MWI artificially obstructs the way to new theories of nature}\label{obstruction}

As we have clarified in detail in the previous subsections, MWI relies on its assumption of universality in order to explain the observed data, hence a single post-quantum event would undermine the whole explanatory power of MWI. Thus, believers of  MWI would likely be more reluctant to accept new evidences begging for an explanation outside of the theory itself. A theory that does not suffer from the fallacy of the HIL, on the other hand, has already a well-established (approximate) explanatory power over the proven finite domain of applicability, without relying on an \emph{a priori} presumed universality; therefore, it does not feel threatened by other theories applying to a larger set of phenomena, or even outperforming it in accuracy for the same phenomena. There is no true conflict between them, rather there must be complementarity and agreement, and in any case they can co-live as the more general theory should include the older one as a limiting case. But, as already argued, any phenomena escaping MWI would make the explanatory power of the whole theory collapse. 

So, the very same tacit presumption of universality that ultimately constitutes its lack of robustness, can be actually used by MWI as a more conservative line of defence of the theory, slowing down the progress of science. Indeed, theories that demarcate their finite range of applicability, or at least \emph{consistently} admit that such range may at some point exist, might find themselves in front of empirical evidence which, they must acknowledge, are incapable of giving an explanation to, and react accordingly just by humbly leaving the corresponding phenomena outside their scope. However, in front of some empirical evidence that MWI may be struggling to give a \emph{concrete} explanation, the tacit universality always provides a joker to play: The theory is considered universally correct so the observed phenomena must bend to it, one way or another.

An example of how adopting MWI can be pernicious to the development of science is provided in~\cite{gao2021time}. Therein, it is stated that observed events that are believed to be anomalous (i.e.\ extremely unlikely in the space of possible configurations)--such us the low entropy of the initial state of the Universe to explain the thermodynamical arrow of time--do not beg for further explanation. MWI can simply accept them by invoking the fact that we happen to be in the decohered branch of the state of the Universe where such events happen. In this way, any event expected to be rare or anomalous does not cry for an explanation in MWI. This is a dangerous way of thinking about scientific investigation itself, because the MWI seems to endorse the view that, since all possible things \emph{do happen} in different worlds, we should blindly accept anything without wonder (see~\cite{baumann2022many} for a criticism). It’s just \emph{per accidens} that we experience to live in this world.

We see in this example how MWI can appeal to the existence of different universes within the theory, in order to automatically embrace any possible experimental evidence within its domains. It is clearly just a rhetorical solution, but it is quite distinctive from a theory entering the HIL: If, consciously or not, we have already assumed that the theory applies to arbitrary phenomena, we do not feel in the need to question whether a given phenomenon can be accounted for with the theory or not. We may rather tend to contort the arguments so that there is a way in which it can be accounted for, especially if otherwise everything we believed in is suddenly lost forever.
But then, asking why things happen the way they do, i.e., asking for a scientific explanation, becomes meaningless, since we are always blinded to possible new explanations.

\section{Discussion and outlook}
\label{outlook}

In this paper we have put forward a novel argument aimed to show why the many-worlds interpretation should be rejected. It ought to be noted that the (numerous) already existing criticisms of MWI arise as conceptual problems ``inside'' the theory, i.e., only once one has accepted the plausibility of the interpretation in the first place. 
Our criticism is here more radical: It shows that the claim of universality is the result of a vicious inference that cannot be justified in the first place. Following a teleological view of unification and universality, the many-worlds interpretation enters the circular argumentation of the HIL, violating--like any other theory entering the HIL--the necessary conditions of tenability for a scientific theory.

We have focused our criticism here on the MWI. It is our view that other interpretations with claims of universality, such as ``superdeterminism'' \cite{hossenfelder2020rethinking} or the ``Lagrangian view'' \cite{wharton2015universe} may also enter the HIL, although the specific reasons could greatly differ. We shall address them in future works.

Furthermore, some arguments we have raised in the present article may have implications for other interpretations that share relevant features with MWI. In particular, we would like to make an important remark on the implications of assuming the quantum nature of the observer of a quantum measurement, something which permeates several interpretations and is pivotal to some discussions on quantum foundations, such as Wigner's friend scenarios \cite{wigner1961remarks}. In principle, the arguments exposed here do not openly confront the legitimacy of such an assumption \emph{per se.} The problem arises when such assumption becomes a necessary ingredient in understanding why the observer does perceive a unique outcome in each measurement, or in explaining the long run statistic of those outcomes. 

Almost a century of success of quantum mechanics has relegated the possibility of its future overcomings to the corner of ideas that one shall gently give a nod to, but then act as if they did not exist. Proposing that such envisioned possible overcomings may have some implications in our current understanding of the theory is felt as an unjustified attack to quantum mechanics. We would like to overturn such view, and consider this work partly as a tribute to quantum mechanics. The unblemished success of the theory in such ample range of phenomena is really staggering. Precisely because of that, it is suicidal to leave our best comprehension of such rounding success in hands of any interpretation that, due to its soaring ambition, is incapable of building itself on any concrete empirical ground, and therefore cannot but fall apart sooner or later.

As we already mentioned, interpretations that admit a Heisenberg cut already own their everlasting position as valid theories in the history of physics, despite all the conceptual problems. 
In contrast, we cannot avoid noticing the sort of irrational desperation that seems to animate certain interpretations, including (but not only) MWI. By trying to ``fill the gap'' left by the admittedly many conceptual problems of Copenhagen Interpretation, they jump to unjustifiable constructions. Whatever unconventional and mind-blowing these constructions may \textit{prima facie} look, far from being revolutionary, they usually pursue a rather conservative attitude in preserving some ``philosophical prejudices" \cite{del2020demolishing} as something necessary and universal. Be them determinism, locality, unitarity, linearity, relativistic causality, etc. We believe that such an attitude can actually be harmful to scientific progress, since it steers in the direction of building up more and more complicated conceptual and mathematical constructions that make the desired ``good idea'' survive at any price. Even at the unacceptable price of ignoring the reality of the scientific process itself! One can uphold strong arguments on why any of those ideas should be kept, and legitimately try to build reasonable interpretations or theories that are coherent with such choice. But none of these arguments can justify playing gods.

In the specific case of MWI, there seems to be an almost religious sentiment that animates its supporters by \textit{believing} that everything that exists is a single, ``simple", immutable, elegant mathematical object, which supposedly lives in an abstract Hilbert space. In this view, everything we observe and experience, including the space in which we move and live, would just be emerging from the only ``real" entity--the universal wave function \cite{carroll2019mad}. With the arguments exposed in this article, we then join Heisenberg here who, to similar claims put forward by Felix Bloch, once simply replied: ``Nonsense, space is blue and birds fly through it."  \cite{bloch1976heisenberg}.

\section*{Acknowledgements} \label{sec:acknowledgements}
We would like to thank Markus M\"uller, Thomas D.\ Galley, Esteban Castro-Ruiz, Veronika Baumann, Philippe Allard Gu\'erin, \v{C}aslav Brukner and Philip Goyal for useful conversations during the elaboration of this paper. F.D.S.\ acknowledges support from FWF (Austrian Science Fund) through an Erwin Schr\"odinger Fellowship (Project J 4699s). L.C.B.\ acknowledges financial support by the Austrian Science Fund (FWF) through BeyondC (F7103-N48). This publication was made possible through the support of the ID 61466 grant from the John Templeton Foundation, as part of The Quantum Information Structure of Spacetime (QISS) Project (qiss.fr). The opinions expressed in this publication are those of the authors and do not necessarily reflect the views of the John Templeton Foundation.

\bibliography{biblio}

\end{document}